\documentclass[aps,twocolumn,noshowpacs]{revtex4}
\usepackage{amsmath,graphicx,amsbsy,amssymb,amsmath,epsfig,latexsym,wasysym,pifont,mathrsfs,natbib}
\usepackage{float} \newfloat{widefig}{thp}{lop}
\usepackage{comment}
\usepackage{array, makecell}
\usepackage{verbatim}
\usepackage{datetime}
\usepackage{multirow,array}
\usepackage{hyperref}

\begin{document}

\title{Site occupancy, composition and magnetic structure dependencies of martensitic transformation in Mn$_{2}$Ni$_{1+x}$Sn$_{1-x}$.}
\author{Ashis Kundu}
\affiliation{Department of Physics, Indian Institute of Technology
  Guwahati, Guwahati-781039, Assam, India.}
\author{Subhradip Ghosh}
\affiliation{Department of Physics,
  Indian Institute of Technology Guwahati, Guwahati-781039, Assam,
  India.}
\date{\today (\currenttime)}

\begin{abstract}
A delicate balance between various factors such as site occupancy, composition and magnetic ordering seems to affect the stability of the martensitic phase in Mn$_{2}$Ni$_{1+x}$Sn$_{1-x}$. Using first-principles DFT calculations, we explore the impacts of each one of these factors on the martensitic stability of this system. Our results on total energies, magnetic moments and electronic structures upon changes in the composition, the magnetic configurations and the site occupancies show that the occupancies at the 4d sites in the Inverse Heusler crystal structure play the most crucial role. The presence of Mn at the 4d sites originally occupied by Sn and it's interaction with the Mn atoms at other sites decide the stability of the martensitic phases. This explains the discrepancy between the experiments and earlier DFT calculations regarding phase stability in Mn$_{2}$NiSn. Our results qualitatively explain the trends observed experimentally with regard to martensitic phase stability and the magnetisations in Ni-excess, Sn-deficient Mn$_{2}$NiSn system. 
\end{abstract}

\maketitle

\section{INTRODUCTION}
In recent years, Ni-Mn-Z based magnetic shape memory alloys (MSMA) have drawn much attention due to their multi-functional properties such as magnetic filed induced strain (MFIS), large magneto-caloric effect, magneto-resistance and exchange bias effect~\cite{UllakkoAPL96,MurrayAPL00,SozinovAPL02,ChmielusNM09,MarcosPRB02,HuPRB01,KrenkePRB07,PasqualePRB05,BiswasAPL05,IngaleJAP09}, which are associated with useful applications such as magneto-mechanical actuator~\cite{UllakkoAPL96,SozinovAPL02}, magnetic refrigeration~\cite{KrenkeNM05,AntoniJPCM09,ZhangSR15,MejiaJAP12} and magnetic recording devices~\cite{ShawAM05} respectively. An extensive research  on Ni-excess Ni-Mn-Z alloys demonstrated that the origin of  these effects are  the coupling between the structural and magnetic phase transitions~\cite{KrenkeNM05,KrenkePRB06,MoyaPRB06,KrenkePRB07,AntoniJPCM09}. The phase transformation in these alloys is driven by the Zeeman energy which is the product of applied magnetic field and the magnetization difference between the structural phases. Since the magnetization in Ni-Mn-Z alloys is mainly contributed by the Mn atoms, the route onward was to explore systems with excess Mn. Researchers were, therefore, inclined to  systematically increase the Mn content in the system, mostly at the expense of Ni, achieving Mn$_{2}$NiZ systems finally\cite{LiuAPL05,LiuPRB06,BarmanPRB08}.

In the family of Mn$_{2}$-bases Heusler alloys, Mn$_{2}$NiGa which shows a martensitic transformation (T$_{m}$=270K) near room temperature and a curie temperature(T$_{c}$=588K) well above room temperature, has been studied extensively\cite{LiuAPL05,LiuPRB06}. The large values of martensitic transformation temperature(T$_{m}$) and Curie temperature(T$_{c}$), in comparison to the prototype and intensely studied Ni$_{2}$MnGa, motivated the researchers to try Mn$_{2}$NiZ alloys with different Z elements. Subsequently, first-principles Density Functional Theory (DFT) calculations were performed on Mn$_{2}$NiZ (Z=Al, In, Sn) systems \cite{LiuPRB06,BarmanPRB08,PaulJAP11}. From the total energy surfaces and insignificant changes in their volumes upon a tetragonal distortion of their high temperature cubic phases, these studies concluded that all of them undergo volume-conserving martensitic transformations at low temperatures and thus could be suitable for shape memory applications. However, there were no experimental evidences to back this claim. The only available experimental investigation on this series of compounds was on Mn$_{2}$NiSn \cite{HelmholdtJLCM87,LakshmiBMS02} which found no evidence of structural transformation in this system. However, Mn$_{2}$NiSn drew attention because of a high value of magnetisation and a Curie temperature of 530 K. The high value of magnetisation was surprising since the DFT calculations \cite{ChakrabartiAPL09,PaulJPCM11,PaulJAP11} and experimental measurements \cite{LiuAPL05,LiuPRB06} revealed that the magnetic moments of Mn$_{2}$NiZ compounds are rather low ($\sim 1 \mu_{B}$), contrary to the expectation; the reason being the anti-parallel coupling of the two Mn atoms.  The high magnetisation value in Mn$_{2}$NiSn was attributed to the presence of Mn-Ni antisites\cite{HelmholdtJLCM87}. Another DFT based investigation \cite{PaulJAP14} subsequently established the correlation between presence of antisites and jump in magnetisation in these systems. Thus, it appeared that there is a significant connection between the site occupancies and magnetism in Mn-excess Mn-Ni-Sn system. 

With a motivation to explore the possibility of a martensitic transformation in Mn-excess Mn-Ni-Sn system, a series of experiments were recently performed by varying the composition ratio of Ni and Sn in Mn$_{2}$NiSn compound. It may be noted that martensitic transformations could be observed in Ni$_{2}$MnSn and Ni$_{2}$MnIn systems only when the ratio of Mn to Sn(In) contents were varied \cite{KrenkePRB05,KrenkePRB06,MoyaPRB06,KanomataJMMM09}. The same trend was observed in the experiments on Mn$_{2}$Ni$_{1+x}$Sn$_{1-x}$ alloys. The alluding martensitic transformation was observed for low Sn content compounds. More importantly, the results indicated intricate relationships between the martensitic stability, composition, atomic ordering and magnetic order. Martensitic transformations were observed only for the compositions $0.56 \leq x \leq 0.92$ \cite{XuanAPL10,CollJTAC10,MaAPL11,MaJAP12,XuanPSS14}. The magnetic states of the compounds in this composition range offered interesting perspective regarding the role of magnetism in martensitic stability. The martensitic transformation temperature, which was having a very high value ($\sim 800$ K) for $x=0.92$ was found to decrease with decreasing $x$ bringing it down below room temperature at $x=0.64$. Giant exchange bias was also observed for systems with $x=0.56-0.68$ \cite{XuanAPL10,MaAPL11,SharmaAPL15} in their martensitic phases indicating that there are competing ferromagnetic and antiferromagnetic domains possibly due to distribution of Mn atoms at different sites. The magnetic structure in the martensitic phase for $x>0.64$ was considered to be Paramagnetic by Ma {\it et al} \cite{MaJAP12}, while ordered magnetic phases were inferred from experimental results for $x \leq 0.64$.  The phase diagram \cite{MaJAP12} suggested that in the range $0.52 \leq x \leq 0.6$, where martensitic and magnetic transitions occur below room temperature, both the austenite and martensite phases are magnetically ordered. The martensite phase was found to be losing stability with the increasing order of the magnetic state finally stabilising a magnetically ordered austenite phase at low temperature for $x < 0.56$. Large changes in the magnetisation across the martensitic transformation for $x=0.6$ hinted at the changes in magnetic order along with changes in the crystal structure leading to magneto-structural effects \cite{MaJAP12}. The particular alloy Mn$_{2}$Ni$_{1.6}$Sn$_{0.4}$ was subsequently investigated by many researchers \cite{TianJAP12,GhoshJAP15,RayEPL15,FichtnerMETALS15,ZhaoINTM16}. In a nutshell, the results demonstrated that the composition, site ordering and the magnetic order are coupled with the stability of the martensitic phase in this system. However, there has been no investigation yet to understand this interrelation from a fundamental point of view. 

In this paper, we make such an attempt with the help of first-principles DFT based methods. We look at the effects of composition, site occupancies and magnetic order on the martensitic stability of Mn$_{2}$Ni$_{1+x}$Sn$_{1-x}$ by changing each one of them in a systematic way. We evolve a qualitative picture of the fundamental physics associated with the various factors influencing the phase stability in this system from the results on energetics, magnetic moments and the electronic structures. We put special emphasis on the system Mn$_{2}$Ni$_{1.6}$Sn$_{0.4}$ since all the factors affecting martensitic stability may be at play for this particular system. The paper is organised as follows: in the section II, computational details are given. The results and discussions are presented in section III followed by conclusions.

\section{Computational Methods}

Electronic structure calculations were performed with spin-polarised density functional theory (DFT) based projector augmented wave (PAW) method as implemented in Vienna Ab-initio Simulation Package (VASP).\cite{PAW94,VASP196,VASP299} For all calculations, we have used Perdew-Burke-Ernzerhof implementation of Generalised Gradient Approximation (GGA) for exchange-correlation functional.\cite{PBEGGA96} An energy cut-off of 450 eV and a Monkhorst-Pack\cite{MP89} $11 \times 11 \times11$ $k$-mesh were used for self consistent calculations. A larger k-mesh of $15 \times 15 \times15$ was used for the calculations of the electronic structures. The convergence criteria for the total energies and the forces on individual atoms were set to 10$^{-6}$ eV and $10^{-2}$ eV/\r{A} respectively.

In order to calculate the magnetic moment and density of states (DOS) for a off-stoichiometric compound, close to experimental composition, we used the multiple scattering Green function formalism as implemented in SPRKKR code \cite{EbertRPP11} by employing the coherent potential approximation (CPA) \cite{CPAPRB67}. For self consistent calculations by the SPRKKR code, the full potential spin polarised scaler relativistic Hamiltonian with angular momentum cut-off $l_{max} = 3$ is used along with a converged $k$-mesh for Brillouin zone integrations. The Green's functions are calculated for 32 complex energy points distributed on a semi-circular contour. The energy convergence criterion was set to 10$^{-6}$ eV for the self-consistency cycles. The experimental lattice constants were used in these calculations for comparison with experiments.

\section{Results and Discussions}

The high temperature austenite phase of stoichiometric Mn$_{2}$NiSn  is Hg$_{2}$CuTi or Inverse Heusler(Space group no 216; $F\bar{4}3m$) with four inequivalent Wyckoff positions(4a, 4b, 4c, 4d). The Mn atoms occupy the 4a(0, 0, 0) and 4c(0.25, 0.25, 0.25) Wyckoff positions. We denote them as MnI and MnII respectively. The Ni and Sn atom occupy the 4b(0.5, 0.5, 0.5) and 4d(0.75, 0.75, 0.75) Wyckoff positions respectively\cite{HelmholdtJLCM87,PaulJPCM11}. The same crystal structure is observed in the experiments on Mn$_{2}$Ni$_{1+x}$Sn$_{1-x}$. However, with Ni-excess, Sn-deficient systems, the occupancies at various sites change. Experimental observations predict that instead of the excess Ni atoms occupying the empty Sn-sites, the excess Ni atoms occupy the MnI sites which are crystallographically equivalent, thus displacing the MnI atoms to the empty Sn sites \cite{XuanAPL10,MaAPL11,WuJPDAP11}. This site-occupancy pattern is consistent with the observed one in Heusler alloys where constituents with higher valence electrons occupy symmetric positions (4a and 4b)\cite{HelmholdtJLCM87,LiuPRB08,KandpalJPD07}. Accordingly, we have modelled the Mn$_{2}$Ni$_{1+x}$Sn$_{1-x}$ system with a 16 atom conventional cubic cell where each Wyckoff position corresponding to the Hg$_{2}$CuTi structure has four sub-lattices.  However, any arbitrary composition can't be modelled with the cell of this size. Only compositions with $x=0,0.25,0.5,0.75,1$ can be modelled. Modelling with any arbitrary value of $x$ within the framework of DFT-PAW formalism as implemented in VASP requires very large supercells which is computationally demanding. Due to this limitation, we have focused on understanding the trends in properties with compositional changes, rather than working in a significantly narrow composition range and look for quantitative accuracy. After fixing the site occupancies in line with the experimental prediction and the empirical rule based on valence electron numbers, we look for the possible magnetic configurations. For Ni$_{2}$Mn$_{1+x}$Z$_{1-x}$ systems, it was found out that the magnetic ground state is dependent on the site preferences of the magnetic atoms \cite{ChunmeiPRB13,ChunmeiPRB12}. For stoichiometric Mn$_{2}$NiSn, the Mn atoms(MnI and MnII) are anti-parallely coupled\cite{PaulJPCM11}. For Mn$_{2}$Ni$_{1+x}$Sn$_{1-x}$ , though the MnI and MnII still couple anti-parallely, the MnII and MnIII (The Mn atoms occupying the 4d positions, originally held by Sn in Mn$_{2}$NiSn)  can be coupled parallel or anti-parallely depending upon their inter-atomic distances. One can find the ground state by comparing energies of the system (with a fixed site occupancy pattern) between the possible magnetic configurations \cite{ChunmeiPRB12}.

\begin{figure}[t]
\centerline{\hfill
\psfig{file=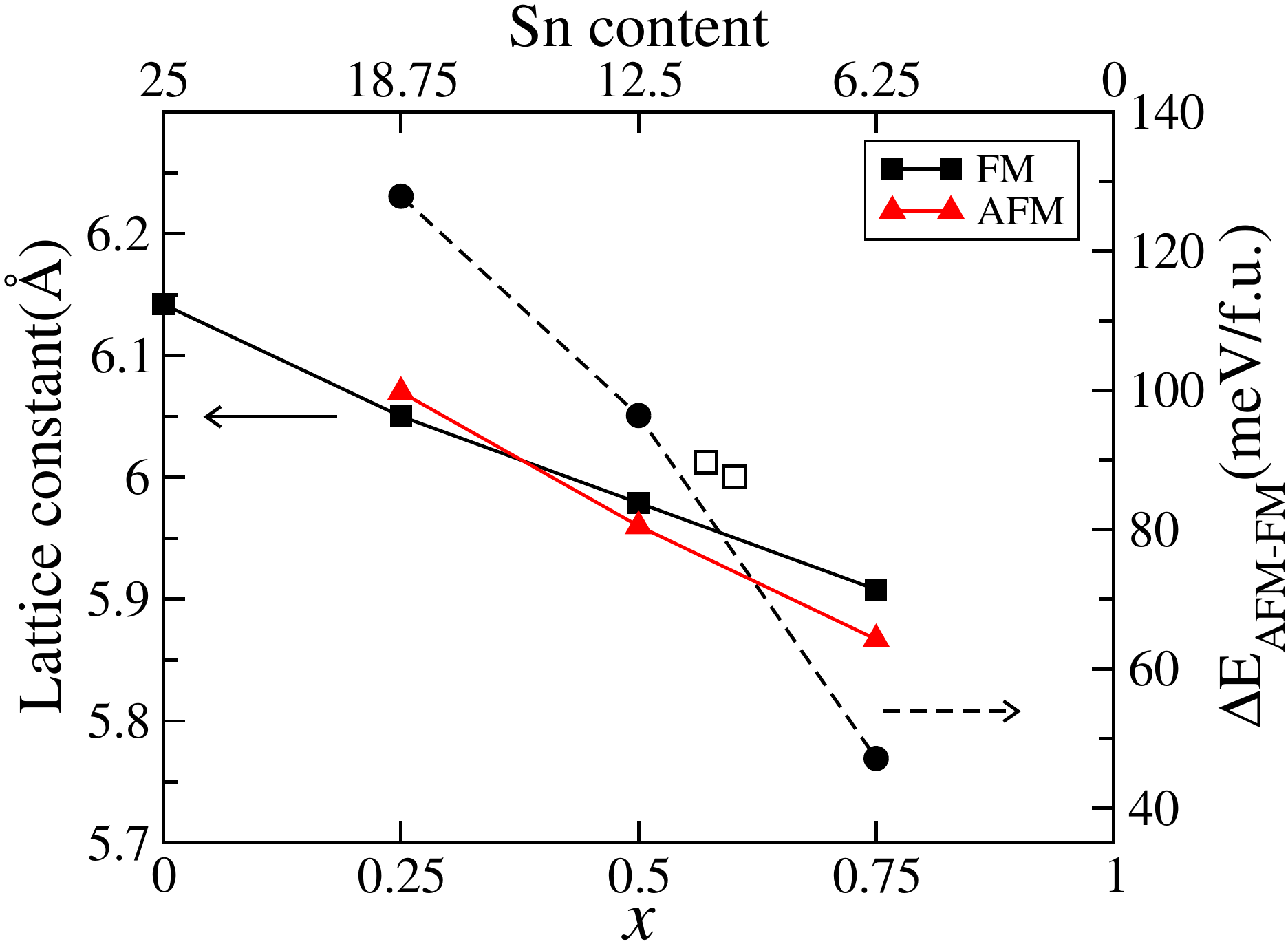,width=0.48\textwidth}\hfill}
\caption{The solid lines correspond to the calculated lattice parameters in the austenite phase of Mn$_{2}$Ni$_{1+x}$Sn$_{1-x}$ with FM and AFM configurations (see text for FM and AFM). The dashed line corresponds to the energy differences between AFM and FM configurations as a function of $x$, in the austenite phase. A positive $\Delta$E indicates the FM configurations have lower energy. Open symbols represent the experimental lattice constants \cite{ZhaoINTM16}. The top $x$-axis represent the Sn content in Mn$_{50}$Ni$_{50-y}$Sn$_{y}$.}
\label{fig1}
\end{figure}

In Fig. \ref{fig1}, the variations in calculated lattice parameters of Mn$_{2}$Ni$_{1+x}$Sn$_{1-x}$ in the austenite phase, as a function of $x$, are shown for both FM(MnII and MnIII align parallel) and AFM(MnII and MnIII align anti-parallel) magnetic configurations. We find that irrespective of the magnetic configuration, the lattice constant decreases with $x$. This can be correlated to the smaller size of Ni (atomic radius 1.35\AA) in comparison with Sn(atomic radius 1.45\AA). Better agreement with the experimental measurements \cite{MaJAP12,ZhaoINTM16} are obtained for the calculated lattice parameters in the FM configuration, as is clear from Fig. \ref{fig1}. To find the ground state magnetic configuration in the austenite phase, we calculated the energy difference($\Delta$E$_{\rm AFM-FM}$) between AFM and FM configurations. The positive value of $\Delta$E$_{\rm AFM-FM}$ for all $x$, indicates that the FM states are energetically favourable(Fig. \ref{fig1}) throughout the composition range in the austenite phase.

\begin{figure}[t]
\centerline{\hfill
\psfig{file=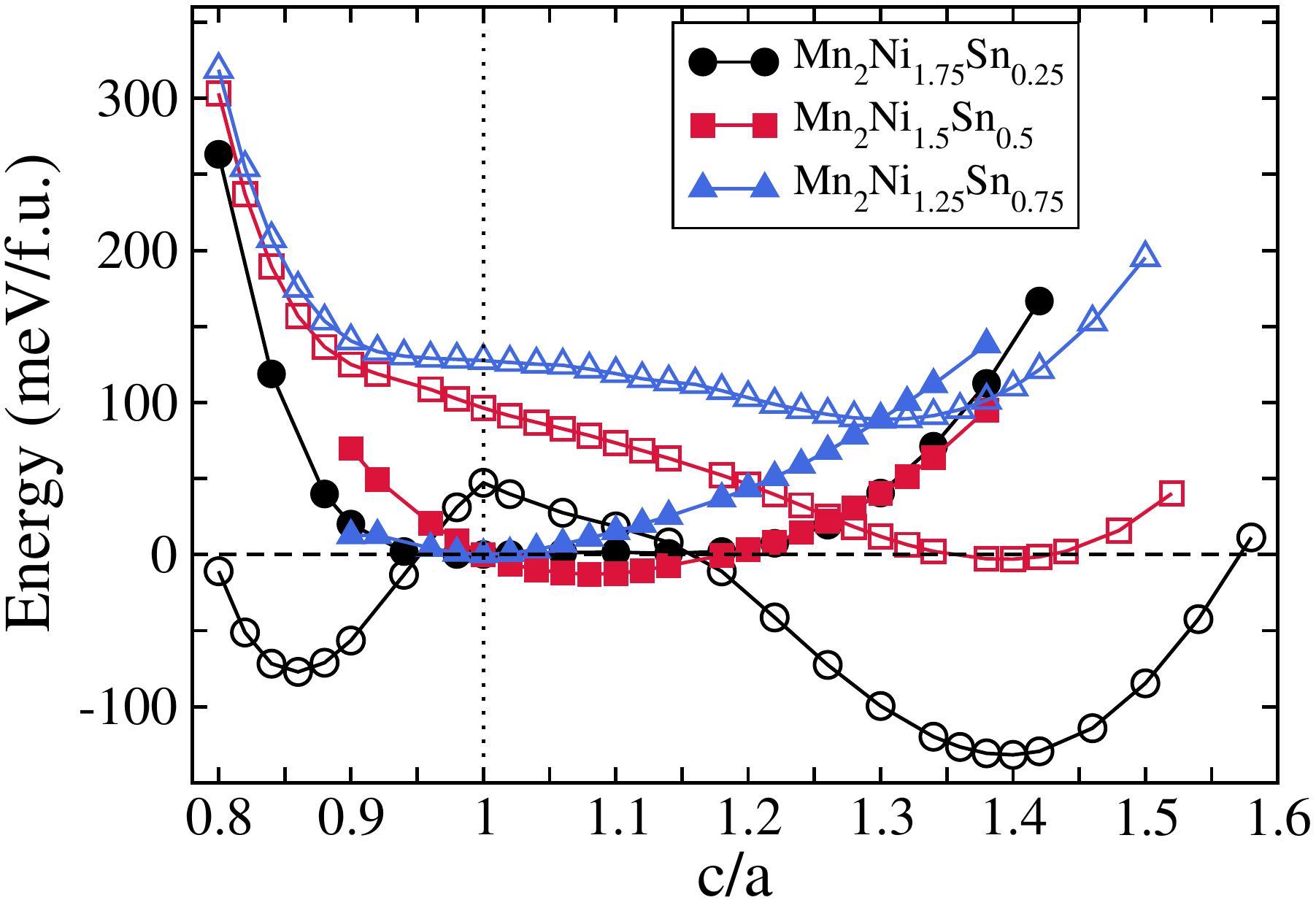,width=0.48\textwidth}\hfill}
\caption{The variation of total energy as a function of tetragonal distortion (c/a) for Mn$_{2}$Ni$_{1+x}$Sn$_{1-x}$ (x=0.75, 0.5, 0.25)  with FM and AFM configurations. The zero energy is the energy of the FM austenite($c/a=1$) phase for all the compositions. Open symbols represent the AFM configurations.}
\label{fig2}
\end{figure}

The martensitic transformation in Mn$_{2}$Ni$_{1+x}$Sn$_{1-x}$ is observed for $x$ between 0.56 and 1. With the cell size used in this work, it is not possible to scan the entire concentration range $x=0.56-1.0$. However, it is possible to investigate the trends in the martensitic transformation by calculating total energies for $x=0.25,0.5,0.75$  after distorting the Hg$_{2}$CuTi structure along $c$-axis. In what follows, we have calculated the total energy as a function of $c/a$ for both FM and AFM configurations. The results are shown in the Fig. \ref{fig2}. In the austenite phase ($c/a=1$), the lowest energy magnetic configuration is FM across compositions. Also, we do not see any energy minima corresponding to a $(c/a)$ in the FM configuration except a shallow one at $x=0.5$. A stable martensitic phase is observed at a $c/a>1$ in this composition but with AFM magnetic configuration. The energy difference between this AFM tetragonal and the FM cubic phase is about 33 meV/atom. For $x=0.5$, the AFM configuration produces a tetragonal ($c/a>1$) phase which is energetically lower than the FM configuration for the system with same $x$, but the energy difference between this tetragonal phase and the global minima in the Hg$_{2}$CuTi phase is only about 1 meV per atom. For the system with $x=0.25$, the energy of the tetragonal ($c/a>1$) phase with AFM configuration is much higher than the global minima in the Hg$_{2}$CuTi phase.  The results, therefore suggest that the martensitic transformation temperature T$_{m}$ will decrease as the Sn content($x$) increases(decreases) since the energy difference between the AFM tetragonal phase and the global FM Hg$_{2}$CuTi phase decreases. For $x=0.75$, the global minima in the tetragonal phase occurs in the AFM configuration and is energetically lower than the global minima in the Hg$_{2}$CuTi phase which has a FM configuration. For $x=0.5$, a composition, almost at the edge of the composition range where martensitic transformation is observed experimentally, the energy of the tetragonal phase is almost same as that of the Hg$_{2}$CuTi global minima. This signifies that the martensitic phase is nearly de-stabilised at this composition. Further increase in Sn content ($x=0.25$) clearly shows that the energy of the tetragonal phase, with AFM configuration, is significantly higher than the energy of the global minima in the Hg$_{2}$CuTi phase obtained with the FM configuration. These results are qualitatively in agreement with the experimental observations \cite{MaJAP12}. Moreover, the following general qualitative picture regarding the effects of site occupancy, composition and the magnetic order on martensitic transformations in Mn$_{2}$Ni$_{1+x}$Sn$_{1-x}$ emerges out of these results: (i) the increase in Sn content at the expense of Ni destabilises the martensitic phase. Ni has played a crucial role in stabilising the martensitic phases in Ni-Mn-Ga systems\cite{ZayakPRB05}. We see the same being repeated in the case of the present system, (ii) the site occupancies are important for the occurrence or disappearance of martensitic phase in this system. Previous DFT calculations on Mn$_{2}$NiSn \cite{PaulJAP11,PaulJAP14} showed a lower total energy in a tetragonal phase ($c/a>1$) than that in the Hg$_{2}$CuTi phase. Inspite of introducing antisites in 4a and 4b sites, it remained so. The results in Fig. \ref{fig2} imply that the energy in the tetragonal phase will remain much higher than the global minimum energy in the cubic phase when $x=0$ that is for the compound Mn$_{2}$NiSn. This will be in agreement with the experimental observations. The key to this will be the site occupancy pattern. In earlier DFT calculations, the 4d site was considered to be fully occupied by Sn. In this work, we see the systematic de-stabilisation of the martensitic phase, evidenced by increasing higher total energy in this phase, when the 4d site is always occupied by Mn atoms. We can, thus, infer that even in Mn$_{2}$NiSn, the 4a and 4d sites will have Mn-Sn binary alloy, (iii) the magnetic configuration in the martensitic phase would be different from that in the austenite phase. The reason behind this can be traced to the distance between the MnII and MnIII atoms. The tetragonal distortion brings these two atoms closer than they were in the austenite phases, introducing antiferromagnetic interaction between them and stabilising the AFM configuration in the martensitic phase.

A physically measurable and important quantity in magnetic compounds that is very sensitive to the site ordering and magnetic structure is the magnetisation. In case of Mn$_{2}$NiZ alloys, earlier DFT calculations have clearly shown this \cite{PaulJPCM11,PaulJAP14}. In case of Mn$_{2}$Ni$_{1.6}$In$_{0.4}$ MSMA, a very high value of saturation magnetisation in the austenite phase was attributed to the co-existence of different magnetic domains-one with usual Mn(4a), Mn(4c), Ni(4b), In-Ni(4d) stacking while the other with Mn-Ni(4a),Mn(4b),Ni(4c), Mn-In(4d) stacking, thus, having both anti-parallel (Mn(4c)-Mn(4a)) and parallel (Mn(4c)-Mn(4d)) alignments among Mn atoms \cite{WuAPL11}. Similarly, in Mn$_{2}$Ni$_{1+x}$Sn$_{1-x}$ systems, the magnetisation can be sensitively dependent upon the site occupancy induced magnetic couplings among Mn atoms. Since the magnetisation measurements have not been done on any of the compounds with $x=0.25,0.5,0.75$, we investigate the influence of site occupancy on magnetic moment for Mn$_{2}$Ni$_{1.6}$Sn$_{0.4}$ whose magnetisation in the austenite phase has been investigated by various groups \cite{XuanJAP10,MaJAP12,ZhaoINTM16,XuanPSS14}. The magnetisation measurements by various groups suggest that it will saturate only at very high field; the measured value at a high field of 13 Tesla is 3.37 $\mu_{B}$ per formula unit \cite{MaJAP12}. Using KKR-CPA method as implemented in the SPRKKR code, we have calculated the magnetic moment per formula unit of Mn$_{2}$Ni$_{1.6}$Sn$_{0.4}$ by varying the occupancies of 4a and 4d sites. The results of total energies and magnetic moments for seven possible configurations are given in Table \ref{table1}. When excess Ni completely occupies the 4d site, the energy of the system is highest and the moment is almost quenched due to compensating anti-aligned moments of MnI and MnII. On the other hand, when excess Ni completely occupies 4a site pushing equal amount of MnI to  4d site, the system has the lowest total energy and the highest magnetic moment. This high magnetic moment is due to the ferromagnetic coupling between MnII and MnIII atoms.
\begin{table}[t]
\centering
\caption{\label{table1} The configurations associated with different site occupancy patterns in the austenite phase of Mn$_{2}$Ni$_{1.6}$Sn$_{0.4}$ (x=0.6), their total energies (E$_{tot}$) in $meV/f.u.$ and total magnetic moment (M$_{s}$) in $\mu_{B}/f.u.$}
\resizebox{0.45\textwidth}{!}{%
\begin{tabular}{c@{\hskip 0.1in} l@{\hskip 0.1in}  l@{\hskip 0.1in}  l@{\hskip 0.1in}  l@{\hskip 0.1in}  c  c }
\hline \hline
\multicolumn{6}{c}{Sites and occupancies}\\
Config. & 4a & 4b & 4c & 4d  & E$_{tot}$  &  M$_{S}$  \\ \hline

C$_{1}$   & Mn & Ni & Mn & Ni$_{0.6}$Sn$_{0.4}$                              & 0        & 0.14 \\
C$_{2}$   & Mn$_{0.9}$Ni$_{0.1}$ & Ni & Mn & Mn$_{0.1}$Ni$_{0.5}$Sn$_{0.4}$  & -70.791 & 1.06 \\
C$_{3}$   & Mn$_{0.8}$Ni$_{0.2}$ & Ni & Mn & Mn$_{0.2}$Ni$_{0.4}$Sn$_{0.4}$  & -165.755 & 1.97 \\
C$_{4}$   & Mn$_{0.7}$Ni$_{0.3}$ & Ni & Mn & Mn$_{0.3}$Ni$_{0.3}$Sn$_{0.4}$  & -285.681 & 2.88 \\
C$_{5}$   & Mn$_{0.6}$Ni$_{0.4}$ & Ni & Mn & Mn$_{0.4}$Ni$_{0.2}$Sn$_{0.4}$  & -428.402 & 3.80 \\
C$_{6}$   & Mn$_{0.5}$Ni$_{0.5}$ & Ni & Mn & Mn$_{0.5}$Ni$_{0.1}$Sn$_{0.4}$  & -590.361 & 4.71 \\
C$_{7}$   & Mn$_{0.4}$Ni$_{0.6}$ & Ni & Mn & Mn$_{0.6}$Sn$_{0.4}$            & -782.417 & 5.60 \\
\hline\hline
\end{tabular}
}
\end{table}

\begin{table*}[t]
\centering
\caption{\label{table2} Possible site occupations of Mn$_{2}$Ni$_{1+x}$Sn$_{1-x}$ ($x=0, 0.25, 0.5, 0.75$) in the 16 atom supercell and their physical properties. a$^{\rm FM}_{0}$ and a$^{\rm AFM}_{0}$ are the equilibrium lattice constants with FM and AFM configurations in the austenite phase respectively. M$^{\rm FM}_{tot}$ is the calculated total magnetic moment with the FM configuration.}
\begin{tabular}{c@{\hskip 0.1in} l@{\hskip 0.1in}  l@{\hskip 0.1in}  l@{\hskip 0.1in}  l@{\hskip 0.1in}  l@{\hskip 0.25in}  c  c  c}
\hline \hline
\multicolumn{6}{c}{Sites and occupancies}\\ \hline
Systems & 4a & 4b & 4c & 4d  & system name  & a$^{\rm FM}_{0}$(\AA) & a$^{\rm AFM}_{0}$(\AA) & M$^{\rm FM}_{tot}$($\mu_{B}/f.u.$) \\ \hline

Mn$_{8}$Ni$_{4}$Sn$_{4}$ & Mn$_{4}$ & Ni$_{4}$ & Mn$_{4}$ & Sn$_{4}$  & Sn4 & 6.142  & - & 0.57  \\
(Mn$_{2}$NiSn) &  &  &  &   &  &     \\  \hline

Mn$_{8}$Ni$_{5}$Sn$_{3}$ & Mn$_{4}$ & Ni$_{4}$ & Mn$_{4}$ & Sn$_{3}$Ni$_{1}$  &  Sn3Ni1 &  6.060 & - & 0.30 \\
(Mn$_{2}$Ni$_{1.25}$Sn$_{0.75}$) & Mn$_{3}$Ni$_{1}$ & Ni$_{4}$ & Mn$_{4}$ & Sn$_{3}$Mn$_{1}$  & Sn3Mn1 & 6.052 & 6.064 & 2.44\\ \hline

Mn$_{8}$Ni$_{6}$Sn$_{2}$ & Mn$_{4}$ & Ni$_{4}$ & Mn$_{4}$ & Sn$_{2}$Ni$_{2}$  & Sn2Ni2 & 5.956 & - & 0.14  \\
(Mn$_{2}$Ni$_{1.5}$Sn$_{0.5}$) & Mn$_{3}$Ni$_{1}$ & Ni$_{4}$ & Mn$_{4}$ & Sn$_{2}$Mn$_{1}$Ni$_{1}$ & Sn2Mn1Ni1 & 5.965 & 5.969 & 2.04 \\
                            & Mn$_{2}$Ni$_{2}$ & Ni$_{4}$ & Mn$_{4}$ & Sn$_{2}$Mn$_{2}$  & Sn2Mn2 & 5.979 & 5.960 & 4.28\\ \hline

Mn$_{8}$Ni$_{7}$Sn$_{1}$ & Mn$_{4}$ & Ni$_{4}$ & Mn$_{4}$ & Sn$_{1}$Ni$_{3}$  & Sn1Ni3 & 5.845 & - & 0.14 \\
(Mn$_{2}$Ni$_{1.75}$Sn$_{0.25}$) & Mn$_{3}$Ni$_{1}$ & Ni$_{4}$ & Mn$_{4}$ & Sn$_{1}$Mn$_{1}$Ni$_{2}$  & Sn1Mn1Ni2 & 5.856 & 5.860 & 2.08 \\
                     & Mn$_{2}$Ni$_{2}$ & Ni$_{4}$ & Mn$_{4}$ & Sn$_{1}$Mn$_{2}$Ni$_{1}$  & Sn1Mn2Ni1 & 5.878 & 5.866 & 4.22\\
                   & Mn$_{1}$Ni$_{3}$ & Ni$_{4}$ & Mn$_{4}$ & Sn$_{1}$Mn$_{3}$  & Sn1Mn3 & 5.908 & 5.867 & 6.44\\                 
\hline \hline
\end{tabular}
\end{table*}

Inspite of having the lowest total energy, the magnetic moment of configuration C$_{7}$ disagrees with the measured saturation magnetisation significantly. On the other hand, the measured value of saturation magnetisation lies in  between the values calculated in the configurations C$_{4}$ and C$_{5}$ which have Ni content of 30$\%$ and 20$\%$ at 4d site respectively. The gradual decrease of magnetic moment with increasing Ni content at 4d site is the result of progressive weakening of MnII-MnIII ferromagnetic component and subsequent strengthening of MnI-MnII antiferromagnetic component. Based upon these results, we conclude that there can be a mixture of domains with different site occupancy patterns in the polycrystalline sample used for magnetic measurements.

\begin{figure}[t]
\centerline{\hfill
\psfig{file=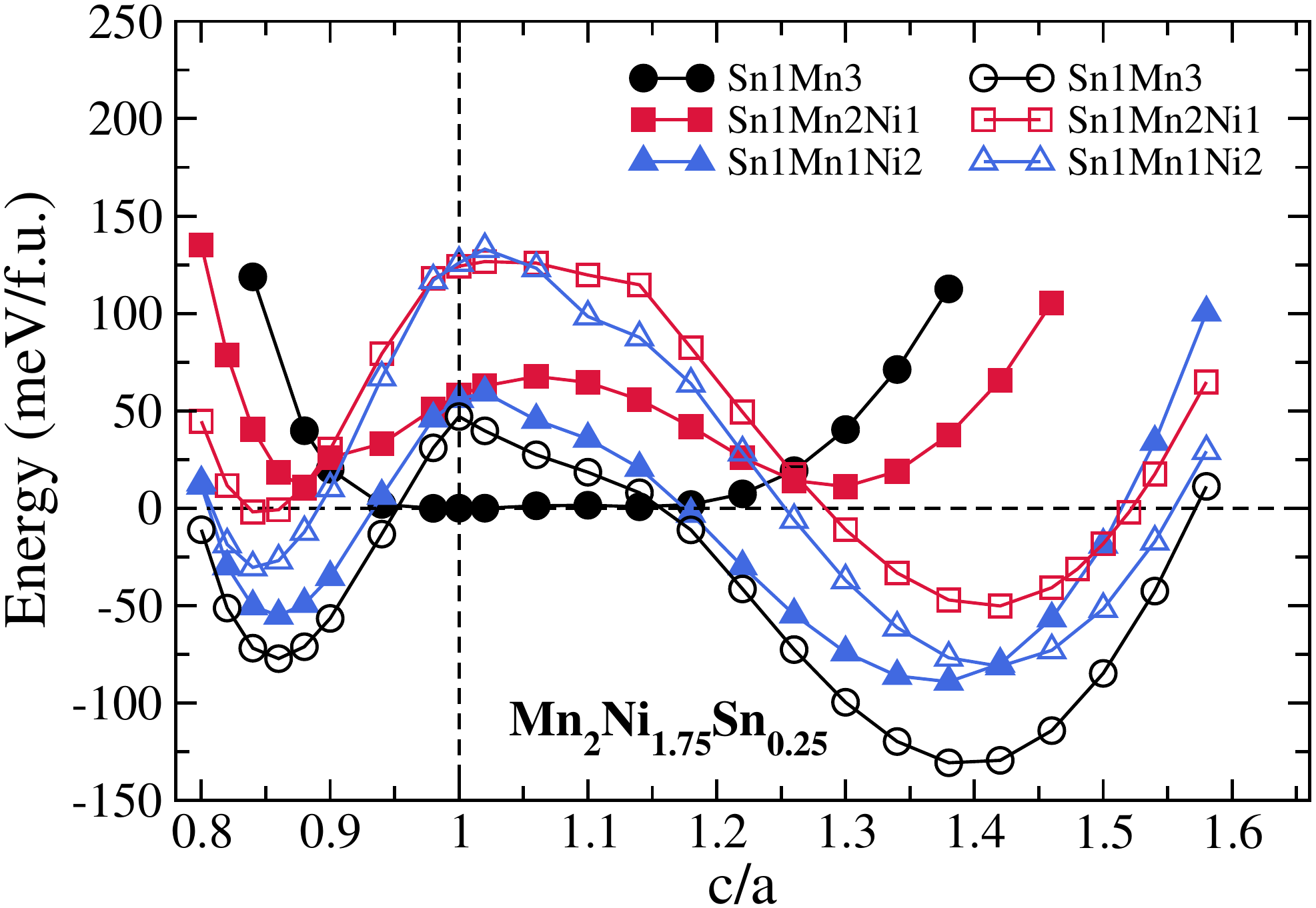,width=0.48\textwidth}\hfill}
\caption{The variation of total energy as a function of c/a ratio for Mn$_{2}$Ni$_{1.75}$Sn$_{0.25}$ with possible configurations given in Table. \ref{table2}. The zero energy is the energy of the Sn1Mn3 system in the FM austenite phase. Open symbols represent the AFM states.}
\label{fig3}
\end{figure}

\begin{figure}[t]
\centerline{\hfill
\psfig{file=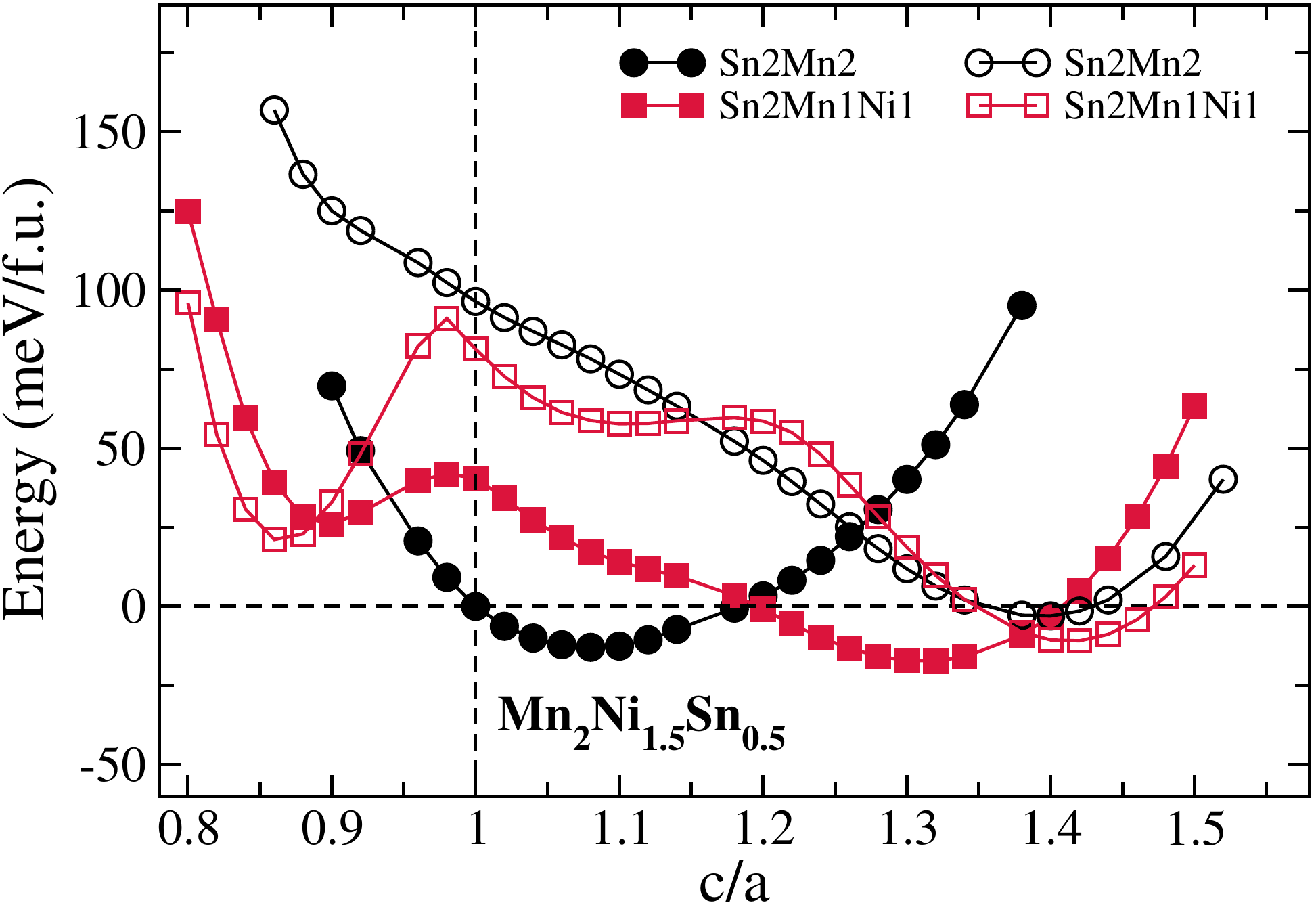,width=0.48\textwidth}\hfill}
\caption{The variation of total energy as a function of c/a ratio for Mn$_{2}$Ni$_{1.5}$Sn$_{0.5}$ with possible configurations given in Table. \ref{table2}. The zero energy is the energy of the Sn2Mn2 system in the FM austenite phase. Open symbols represent the AFM states.}
\label{fig4}
\end{figure}

The question now remains is whether site occupancy patterns with some Ni at the 4d site can reproduce the trend with regard to martensitic stability as a function of composition as observed experimentally and is explained through Fig. \ref{fig2}. To understand the effect of site occupancy on the stability of the martensitic phases, we have calculated the total energies as a function of  $c/a$ for Mn$_{2}$Ni$_{1+x}$Sn$_{1-x}$($x=0.5,0.75$) with different occupancies of the 4a and 4d sites for each of FM and AFM magnetic configurations. The results are shown in Fig. \ref{fig3} and \ref{fig4}. The configurations considered for each $x$ are given in Table \ref{table2}. The system name against each site occupancy pattern is given in column 6 of Table \ref{table2}. In the results presented in Fig. \ref{fig3} and \ref{fig4}, we have not included the cases where no Mn atom occupies the 4d sites (systems Sn1Ni3 and Sn2Ni2). This is because the magnetic moments calculated in these configurations are extremely low (Table \ref{table2}), clearly in severe disagreement with the experiments as is seen in case of $x=0.6$ (Table \ref{table1}). Moreover, for any $x$, such configurations always produced lowest total energies for a $c/a \neq 1$, indicating a energetically favourable martensitic phase at low temperature irrespective of the Ni(Sn) content in the system which is in contradiction to the experimental observations. 
\begin{figure}[t]
\centerline{\hfill
\psfig{file=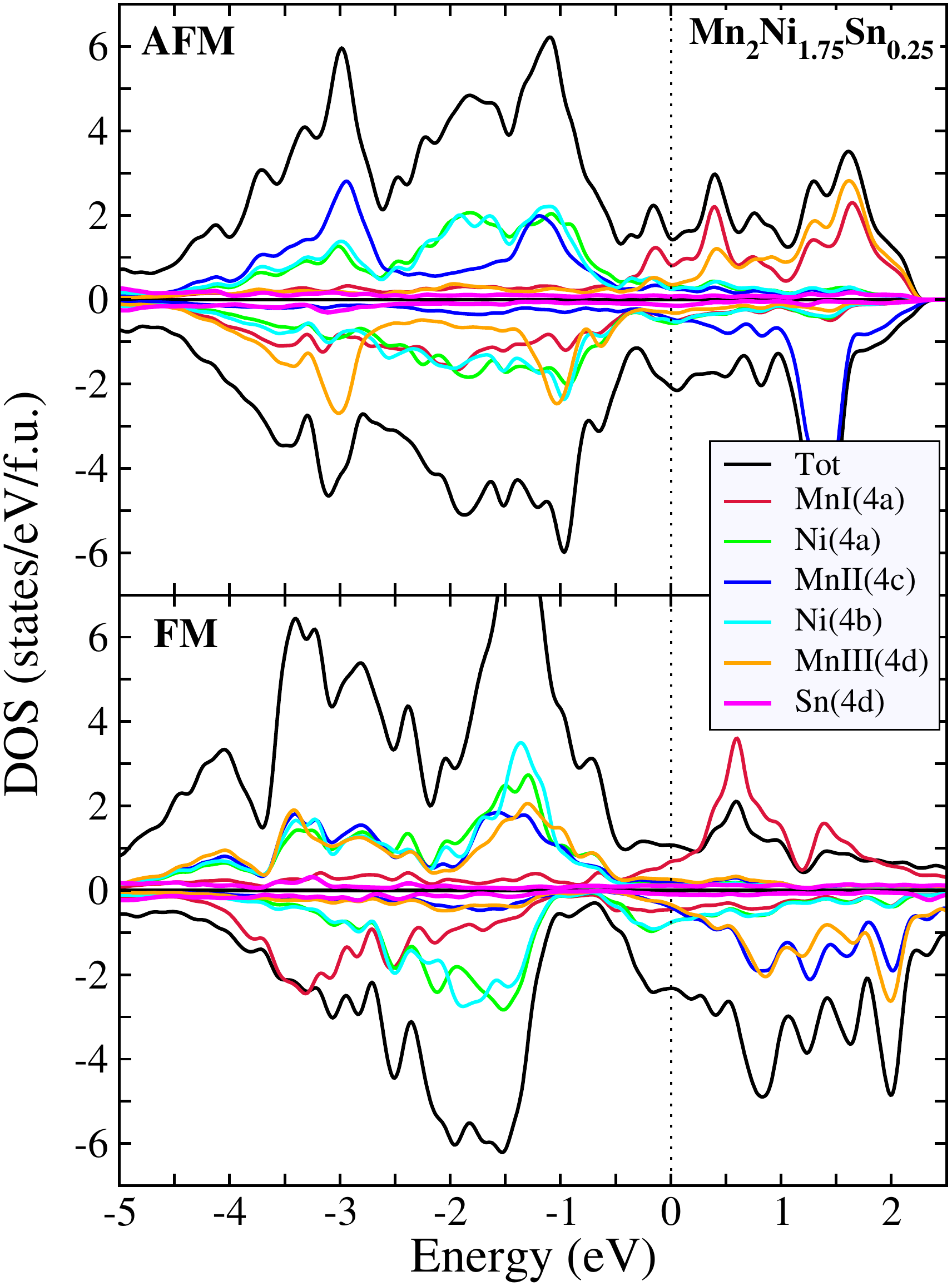,width=0.45\textwidth}\hfill}
\caption{Total and atom projected density of states of Mn$_{2}$Ni$_{1.75}$Sn$_{0.25}$ (Sn1Mn3) system with FM and AFM configurations in the austenite phase. The zero of the energy is set at Fermi level(E$_{\rm F}$).}
\label{fig5}
\end{figure}

In case of Mn$_{2}$Ni$_{1.75}$Sn$_{0.25}$, the FM magnetic structure provides the lowest total energy for the austenite phase ($c/a=1$) when the 4d site has Mn and Sn only (the system Sn1Mn3). The total energy curve does not imply any possibility of a martensitic transformation with FM structure in this system. When a fraction of Ni atoms starts occupying the Sn site (Sn1Mn2Ni1 and Sn1Mn1Ni2), a tetragonal phase produces lower total energy in the FM structure indicating a martensitic transformation(Fig. \ref{fig3}). When 4d site has 50$\%$ of Ni, the tetragonal phase with $c/a>1$ is energetically lower than the global minimum for the austenite phase ($c/a=1$). With the AFM magnetic structure, all site occupancies stabilise the martensitic phases as can be understood from total energies being lower in the tetragonal phases than the global minimum in the austenite phase. Two things, therefore, emerge: (i) at this composition, the anti-ferromagnetic interaction between MnII and MnIII atoms stabilise the martensitic phase, irrespective of the site occupancies at the 4a and 4d sites, and (ii) the presence(absence) of more Ni(Mn) at the 4d site stabilises the martensitic phase irrespective of the magnetic structure. Thus we see comparable total energies  in the FM and AFM configurations for the tetragonal phases of the system Sn1Mn1Ni2( 50$\%$ of Ni at 4d site) near their respective energy minima.  For Mn$_{2}$Ni$_{1.5}$Sn$_{0.5}$, irrespective of whether Ni occupies 4d site or not, shallow minima at a $c/a \neq 1$ are obtained in the FM configuration. For the AFM configurations, although the total energies in the tetragonal phases are much lower than those in the respective cubic phases, they are comparable to the global minimum in the austenite phase. This implies that at this composition, the martensitic stability is almost insensitive to the site occupancy as long as Mn occupies 4a, 4b and 4d sites. In fact, the shallow minima of the tetragonal structure for all magnetic configurations and all site occupancies suggest that the system has a tendency towards stabilising the austenite phase at the low temperature; the magnetic configuration being the FM and the best stability is achieved for 4d site having Mn and Sn only.

\begin{figure}[t]
\centerline{\hfill
\psfig{file=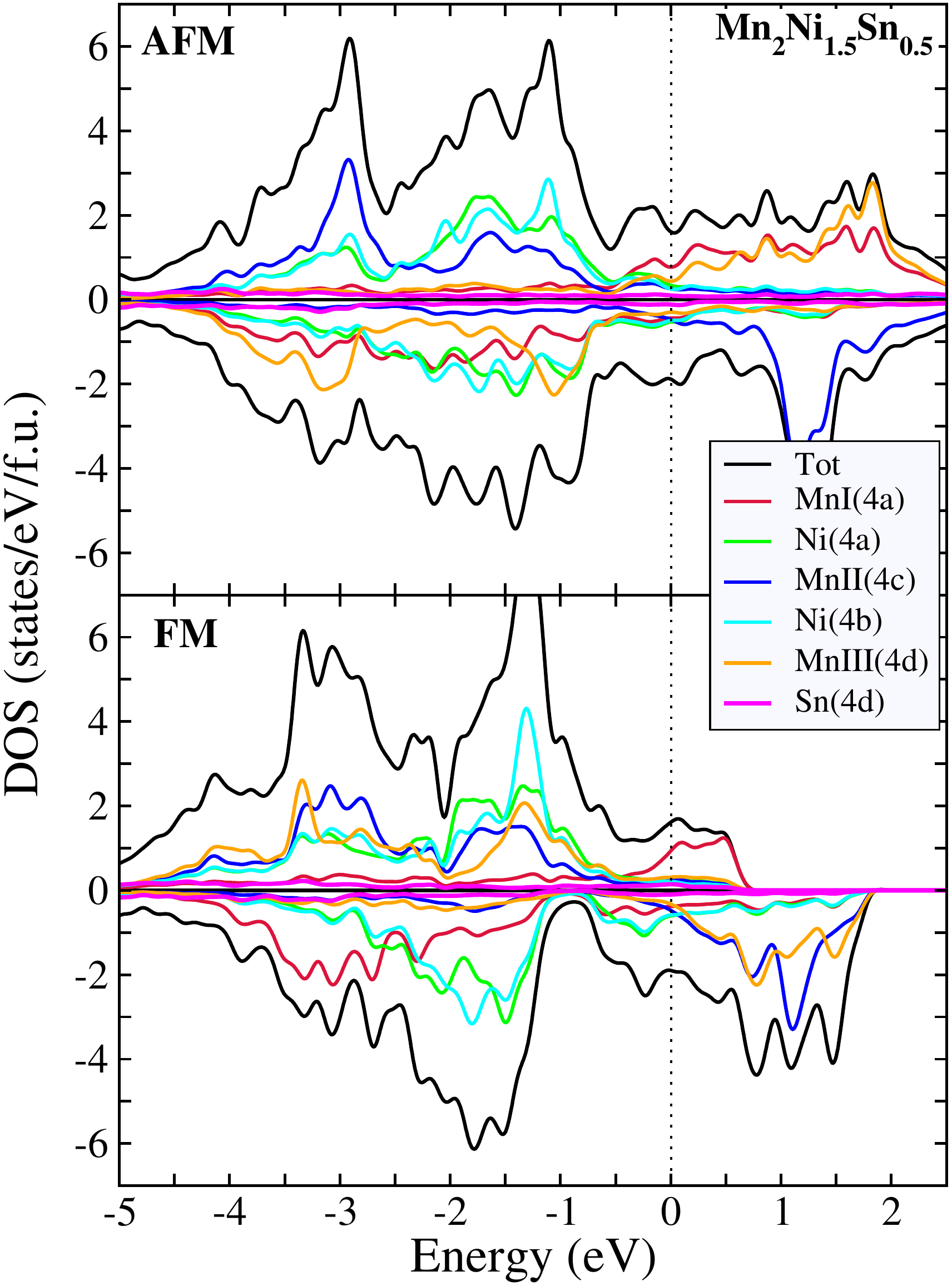,width=0.45\textwidth}\hfill}
\caption{Total and atom projected density of states of Mn$_{2}$Ni$_{1.5}$Sn$_{0.5}$ (Sn2Mn2) system with FM and AFM configurations in the austenite phase. The zero of the energy is set at Fermi level(E$_{\rm F}$).}
\label{fig6}
\end{figure}


In order to understand the connection between the preferred magnetic state and the stability of the martensitic phase from a microscopic point of view, we present the densities of states(DOS) of austenite phases of Mn$_{2}$Ni$_{1.75}$Sn$_{0.25}$ and Mn$_{2}$Ni$_{1.5}$Sn$_{0.5}$, with FM and AFM configurations. The results are shown in Fig. \ref{fig5} and \ref{fig6}, respectively. The densities of states are calculated with Sn1Mn3 and Sn2Mn2 systems(Table \ref{table2}) for Mn$_{2}$Ni$_{1.75}$Sn$_{0.25}$ and Mn$_{2}$Ni$_{1.5}$Sn$_{0.5}$, respectively. In case of FM configuration of Mn$_{2}$Ni$_{1.75}$Sn$_{0.25}$ (Fig. \ref{fig5}), a pseudo-gap is observed at -0.75 eV in the minority DOS which results from hybridisation between Sn-$p$ and Ni-, MnI-$d$ states. This pseudo-gap indicates a strong covalent bonding in the FM states\cite{GelattPRB83,HuPRB03,ChunmeiPRB12} resulting in the stability of the austenite structure with FM magnetic configuration. In case of AFM magnetic configuration, the pseudo-gap disappears due to the additional states of MnIII in the minority DOS around the same energy. In this case, the $d-d$ hybridisation becomes stronger resulting in weaker covalent bonding as compared to the FM case. Thus the austenite phase with AFM configuration has less stability. This corroborates the trends in the energetics presented in Fig. \ref{fig2}.  Another noteworthy feature is that in the AFM configuration, the Fermi energy cuts through a small peak in the minority DOS. This peak is associated with the Jahn-Teller distortion leading to the stability of the martensitic phase. The competitive effect of the strong covalent bonding in the occupied part of the minority DOS giving rise to a pseudo-gap  and the Jahn-Teller effect near the Fermi level giving rise to a peak very near or at the Fermi level has been used to explain the martensitic stability in a number of Ni-Mn-Z systems \cite{ChunmeiPRB13,ChunmeiPRB12}. In the AFM configuration of the present case, the simultaneous disappearing of the pseudo-gap and emergence of the Jahn-Teller peak at the Fermi level explains the stability of the martensitic phase with this magnetic configuration. For Mn$_{2}$Ni$_{1.5}$Sn$_{0.5}$, the pseudo-gap in the minority DOS for FM configuration is deeper indicating a stronger covalent bonding leading to a greater stability of the austenite phase with FM configuration at low temperature in comparison to Mn$_{2}$Ni$_{1.75}$Sn$_{0.25}$. For the AFM magnetic configuration, we once again see the filling of the pseudo-gap due to MnIII states and simultaneous weakening of the covalent bonding giving way to the Jahn-Teller effect. Thus, the electronic structures demonstrate the correlation between the stability of a particular phase with the magnetic configuration. The MnIII states clearly play the decisive role in all the compounds. 
\begin{figure}[t]
\centerline{\hfill
\psfig{file=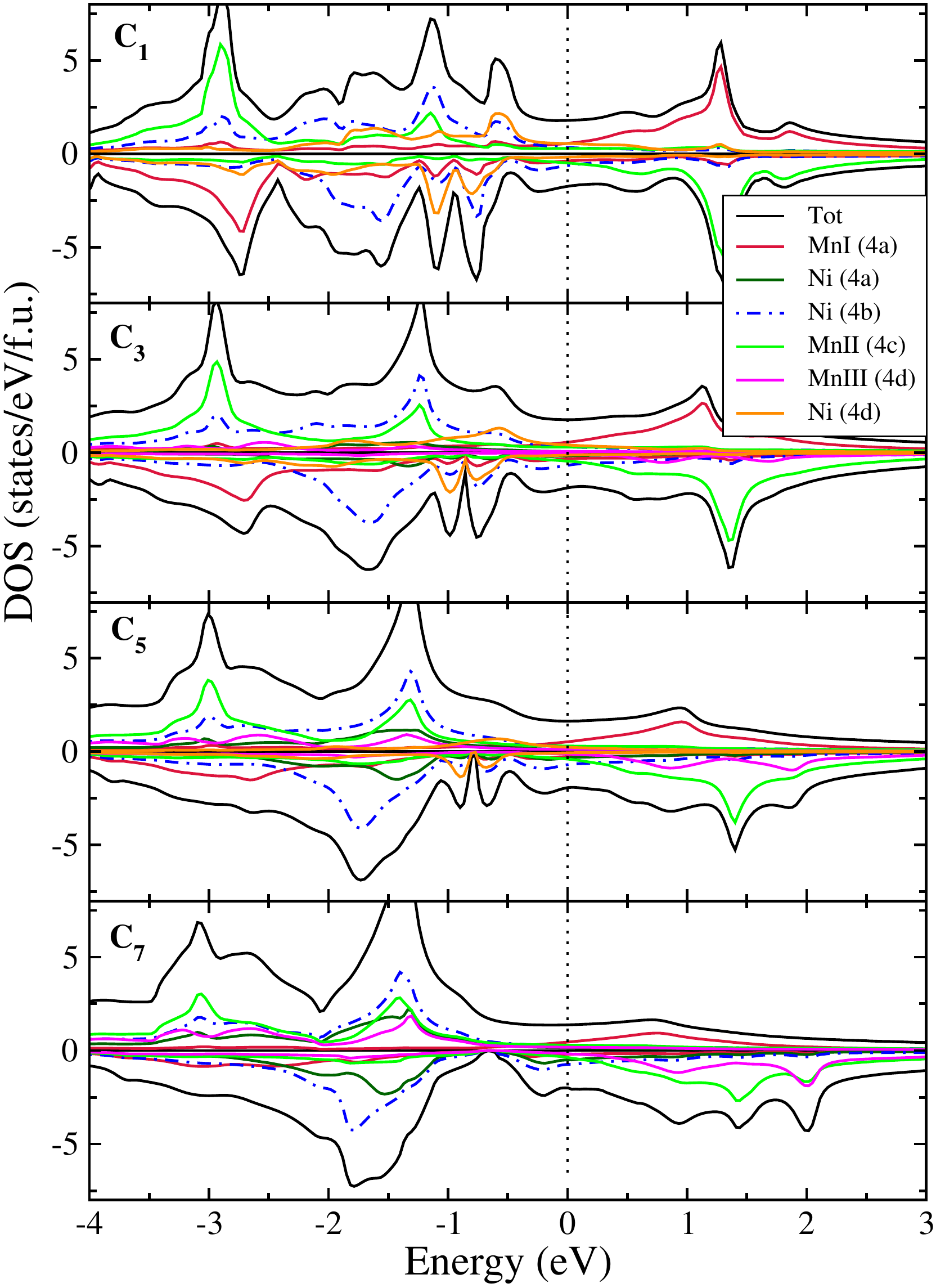,width=0.48\textwidth}\hfill}
\caption{Total and atom projected density of states of Mn$_{2}$Ni$_{1.6}$Sn$_{0.4}$ system with FM configuration in the austenite phase and with different site occupations as presented in Table \ref{table1}. The zero of the energy is set at Fermi level(E$_{\rm F}$).}
\label{fig7}
\end{figure}

After understanding the role of MnIII atoms in stabilising the austenite phase in the FM magnetic configuration, we attempt fundamental understanding of the role of site ordering and stability by analysing the changes in the electronic structures with changes in the composition at the 4d site. For this purpose, we have calculated the DOS of Mn$_{2}$Ni$_{1.6}$Sn$_{0.4}$ as several configurations with both Ni and Mn at the 4d sites had been considered for this system, making it suitable to observe the changes in a finer range. The results with four different configurations taken from Table \ref{table1} are shown in Fig. \ref{fig7}. All calculations are done in the austenite phase with FM configuration. We find significant changes in the minority DOS as more and more Ni occupy 4d sites. The pseudo-gap near -0.75 eV which is prominent in the configuration C$_{7}$ (No Ni in 4d site) starts getting narrower and shallower as more and more Ni occupies 4d sites, due to the states of these Ni. Finally when the 4d site contains only Ni and Sn, the pseudo-gap almost disappears. The gradual weakening of the covalent bond manifested in  disappearance of the pseudo-gap destabilises the cubic phase with FM configuration. The decrease in the magnetic moment from configurations C$_{7}$ to C$_{1}$ (Table \ref{table1}) can be understood from these features as well. The gradual reduction of the MnIII atoms from the system weakens the ferromagnetic part in the magnetic interaction which was mostly due to MnII and MnIII atoms, resulting in the loss of total moment with increase of Ni at the 4d sites. These features provide explanations for the trends in the total energies and magnetic moments given in Table \ref{table1}.

\section{Conclusions}

Using first principles density functional based calculations, we have investigated the interrelations between composition, site occupancy, magnetic structure and martensitic stability in  Mn$_{2}$Ni$_{1+x}$Sn$_{1-x}$ systems and interpreted the results from their electronic structures. We infer that the trends in the martensitic stability as a function of composition, observed experimentally, depends on the occupancy of Mn atoms at the Sn sites. The Ferromagnetic interaction among the Mn atoms at the 4c sites and at the 4d sites stabilise the austenite phases while the antiferromagnetic interactions among them due to shortening of their bond distances under a tetragonal distortion stabilise the martensitic phases. The covalent bonding between the MnI, Ni and Sn minority states stabilise the austenite phases when MnII-MnIII interactions are ferromagnetic while the presence of MnIII states in the minority bands when MnII-MnIII anti-align reduce this stability. Subsequently, the anti-alignment of MnII and MnIII atoms strengthen the competing Jahn-Teller effect leading to the stability of the martensitic phases. Apart from the MnIII atoms, the Ni atoms at the 4d sites also play significant roles in the stability of the martensitic phases in this system. The Ni atoms at the 4d sites, unlike the Ni atoms at the 4b sites, weaken the covalent bonding by filling the pseudo-gap in the minority bands, de-stabilising the austenite phase. Thus, a delicate balance of Ni and Mn atoms in the 4d sites decides the magnetism and the phase stability of Mn$_{2}$Ni$_{1+x}$Sn$_{1-x}$ system. An important  by product of this result is the understanding of the discrepancies between the results obtained from previous DFT calculations and the experimental observations regarding the stability of the martensite phase in Mn$_{2}$NiSn. While the DFT calculations predicted a stable martensite phase at low temperature, the experiments found the austenite phase to be stable down to very low temperature. The trends obtained from the energetics and the electronic structures in the present study indicate that the presence of Mn at the 4d site must have been there in the experimental samples which induces a ferromagnetic coupling between Mn at 4c and Mn at 4d sites leading to the stability of a austenite phase. The results presented in this work, therefore, interpret the experimental observations by pinpointing the roles of a number of factors like the site ordering, the composition and the magnetic interactions, which has not been done yet from a microscopic point of view. This study also provides us with a wider scope of interpreting the experimental results for similar systems like Mn-Ni-In where such interesting interrelations between a number of factors seem to be existing.

\section*{ACKNOWLEDGMENT}
AK thank Arup Ghosh for useful discussion at the beginning at the work. The authors would like to thank IIT Guwahati and DST India for the PARAM Supercomputing facility and the computer cluster in the Department of Physics, IIT Guwahati

\bibliographystyle{aip} 

\end{document}